\documentclass{amsart}
\usepackage{amsmath}
\usepackage{amsthm}
\usepackage{amscd}
\usepackage{amsfonts}
\usepackage{amssymb}
\usepackage{epic}
\usepackage{eepic}

\newtheorem{thm}{Theorem}

\theoremstyle{definition}

\theoremstyle{remark}
\newtheorem{rem}{Remark}
\newtheorem{conj}{Conjecture}

\newcommand{\beq}{\begin{equation}}
\newcommand{\eeq}{\end{equation}}

\newcommand{\id}{\mbox{id}}

\newcommand{\pa}{\partial}
\newcommand{\ot}{\otimes}
\newcommand{\ra}{\rightarrow}

\newcommand{\ti}{\times}

\newcommand{\fr}[2]{{\textstyle \frac{#1}{#2} }}

\newcommand{\fsl}{{\mathfrak s}{\mathfrak l}}

\newcommand{\s}{\text{s}}
\newcommand{\bra}{\langle}
\newcommand{\ket}{\rangle}
\newcommand{\al}{\alpha}
\newcommand{\bal}{\bar{\alpha}}
\newcommand{\be}{\beta}
\newcommand{\ga}{\gamma}

\newcommand{\Ga}{\Gamma}
\newcommand{\de}{\delta}
\newcommand{\De}{\Delta}
\newcommand{\ep}{\epsilon}

\newcommand{\om}{\omega}

\newcommand{\si}{\sigma}
\newcommand{\up}{\Upsilon}

\newcommand{\tE}{\tilde{E}}
\newcommand{\tF}{\tilde{F}}
\newcommand{\tK}{\tilde{K}}

\newcommand{\tU}{\tilde{U}}

\newcommand{\CA}{{\mathcal A}}

\newcommand{\CC}{{\mathcal C}}

\newcommand{\CF}{{\mathcal F}}
\newcommand{\CG}{{\mathcal G}}
\newcommand{\CH}{{\mathcal H}}

\newcommand{\CO}{{\mathcal O}}  
\newcommand{\CP}{{\mathcal P}}

\newcommand{\CS}{{\mathcal S}}

\newcommand{\CU}{{\mathcal U}}
\newcommand{\CV}{{\mathcal V}}

\newcommand{\fx}{{\mathfrak x}}

\newcommand{\fz}{{\mathfrak z}}
\newcommand{\FB}{{\mathfrak B}}

\newcommand{\BR}{{\mathbb R}}

\newcommand{\BC}{{\mathbb C}}

\newcommand{\BS}{{\mathbb S}}

\newcommand{\USL}{\CU_q(\fsl(2,\BR))}

\renewcommand{\Re}{\text{Re}}
\renewcommand{\Im}{\text{Im}}

\def\ew{\hspace*{-1mm}}   \def\ppe{\hspace*{-2.5mm}}

\newcommand{\Fus}[6]{F_{{\scriptstyle #1}{\scriptstyle #2}}
  \hspace*{.3mm}\displaystyle{[} \ew \begin{array}{ll} {\scriptstyle #3 }
  \ppe & {\scriptstyle #4} \ppe \\[-2mm] {\scriptstyle #5}\ppe &
  {\scriptstyle #6}\ew \end{array}\displaystyle{]}}
\newcommand{\CBls}[6]{{#6}_{{\scriptstyle #1}}^s
  \hspace*{.3mm}\displaystyle{[} \ew \begin{array}{ll} {\scriptstyle #2 }
  \ppe & {\scriptstyle #3} \ppe \\[-2mm] {\scriptstyle #4}\ppe &
  {\scriptstyle #5}\ew \end{array}\displaystyle{]}}
\newcommand{\CBlt}[6]{{#6}_{{\scriptstyle #1}}^t
  \hspace*{.3mm}\displaystyle{[} \ew \begin{array}{ll} {\scriptstyle #2 }
  \ppe & {\scriptstyle #3} \ppe \\[-2mm] {\scriptstyle #4}\ppe &
  {\scriptstyle #5}\ew \end{array}\displaystyle{]}}
\newcommand{\CfBls}[5]{\CBls{#1}{#2}{#3}{#4}{#5}{\CF}}
\newcommand{\CfBlt}[5]{\CBlt{#1}{#2}{#3}{#4}{#5}{\CF}}
\newcommand{\OM}[3]{\Omega
  \hspace*{.3mm}\bigl( \ew \begin{array}{c} {\scriptstyle #1 }
  \ppe \\[-2mm] {\scriptstyle #2}\;{\scriptstyle #3}\ew 
\end{array}\bigr)}
\newcommand{\CpBls}[5]{\Phi_{{\scriptstyle #1}}^s
  \hspace*{.3mm}\displaystyle{[} \ew \begin{array}{ll} {\scriptstyle #2 }
  \ppe & {\scriptstyle #3} \ppe \\[-2mm] {\scriptstyle #4}\ppe &
  {\scriptstyle #5}\ew \end{array}\displaystyle{]}}
\newcommand{\CpBlt}[5]{\Phi_{{\scriptstyle #1}}^t
  \hspace*{.3mm}\displaystyle{[} \ew \begin{array}{ll} {\scriptstyle #2 }
  \ppe & {\scriptstyle #3} \ppe \\[-2mm] {\scriptstyle #4}\ppe &
  {\scriptstyle #5}\ew \end{array}\displaystyle{]}}
\newcommand{\CgBls}[5]{\CG_{{\scriptstyle #1}}^s
  \hspace*{.3mm}\displaystyle{[} \ew \begin{array}{ll} {\scriptstyle #2 }
  \ppe & {\scriptstyle #3} \ppe \\[-2mm] {\scriptstyle #4}\ppe &
  {\scriptstyle #5}\ew \end{array}\displaystyle{]}}
\newcommand{\CgBlt}[5]{\CG_{{\scriptstyle #1}}^t
  \hspace*{.3mm}\displaystyle{[} \ew \begin{array}{ll} {\scriptstyle #2 }
  \ppe & {\scriptstyle #3} \ppe \\[-2mm] {\scriptstyle #4}\ppe &
  {\scriptstyle #5}\ew \end{array}\displaystyle{]}}
\newcommand{\CGC}[6]{\displaystyle{[} \,\ew \begin{array}{lll} 
  {\scriptstyle #1} \ppe
  & {\scriptstyle #3} \ppe & {\scriptstyle #5} \ew \\[-2mm] {\scriptstyle
  #2} \ppe & {\scriptstyle #4}\ppe & {\scriptstyle #6} \ew\end{array}
  \displaystyle{]}}
\newcommand{\SJS}[6]{ \displaystyle{\bigl\{ } \ew 
\begin{array}{ll} {\scriptstyle #1 }
  \ppe & {\scriptstyle #2} \ppe \\[-2mm] {\scriptstyle #3}\ppe &
  {\scriptstyle #4}\ew \end{array}\big| \ew
\begin{array}{l} {\scriptstyle #5 }
  \ppe \\[-2mm] {\scriptstyle #6}\ew  \end{array}\displaystyle{\bigr\}_b}} 
\newcommand{\SJSL}[6]{ G_{{\scriptstyle #5}{\scriptstyle #6}}
  \hspace*{.3mm}\displaystyle{[} \ew \begin{array}{ll} {\scriptstyle #3 }
  \ppe & {\scriptstyle #2} \ppe \\[-2mm] {\scriptstyle #4}\ppe &
  {\scriptstyle #1}\ew \end{array}\displaystyle{]} } 
\newcommand{\rf}[1]{(\ref{#1})}

\newcommand{\aufz}
{\begin{list}{$\bullet$}{\topsep0cm \itemsep0cm \parsep0cm}}
\newcommand{\eaufz}{\end{list}}
\newcounter{num}
\newcommand{\remlst}{\begin{list}
{(\arabic{num})}{\usecounter{num}\topsep0cm \itemsep0cm \parsep0cm}}
\begin{document}
\thispagestyle{empty}
\hspace*{\fill} \begin{minipage}{3cm} DIAS-STP-99-14\\
ESI-791\\
LPM-99/46\end{minipage}\\[.5cm]
\title{Liouville bootstrap via harmonic analysis on a noncompact 
quantum group}
\author{\sc B. Ponsot, J. Teschner}
\address{B.P.: Laboratoire de Physique Math\'ematique,
Universit\'{e} Montpellier II,
Pl. E. Bataillon, 34095 Montpellier, France}
\email{ponsot@lpm.univ-montp2.fr}
\address{J.T.: School for theoretical Physics, Dublin Institute for Advanced
Studies, 10 Burlington Road, Dublin 4, Ireland}
\email{teschner@stp.dias.ie}

\begin{abstract}
The purpose of this short note is to announce results that amount to 
a verification of the bootstrap for Liouville theory in the 
generic case under certain assumptions concerning existence and properties
of fusion transformations. 
Under these assumptions one may characterize the fusion and braiding 
coefficients as solutions of a system of functional equations that follows
from the combination of consistency requirements and known results. 
This system of equations has a unique solution for irrational central charge
$c>25$. The solution is 
constructed by solving the Clebsch-Gordan problem for a certain
continuous series of quantum group representations and constructing
the associated Racah-coefficients. This gives an 
explicit expression for the fusion coefficients. 
Moreover, the expressions can be continued into the strong coupling 
region $1<c<25$, providing a solution of the bootstrap also for this region.
\end{abstract}

\maketitle
\section{Introduction}
Liouville theory or close relatives of it such as the $H_3^+$ or 
$SL(2)/U(1)$ WZNW models play a central role in a variety of string-theoretical
or gravitational models. These models are simple enough to justify
the hope for exact results yet rich enough to capture some important
aspects of the physics of strings on nontrivial (maybe curved) backgrounds.
From another point of view, these are the natural starting points for 
beginning to investigate {\it noncompact} conformal field theories (CFT),
i.e. CFT with a continuous spectrum of primary fields.
This has motivated a lot of effort towards the exact solution of these 
models.

Both from the point of view of applications and of
the intrinsic structure of the CFT one may consider the determination
of the spectrum Virasoro-representations 
and of the three point function to be central objectives.
Concerning the former, a reasonable conjecture was obtained in the 
early work \cite{CT} (stated in \rf{Liouspec} below). 
As far as the three point function is concerned, important progress was
initiated by the papers \cite{DO,ZZ} where an explicit formula was proposed
and checked in various ways. A method to derive this formula from 
conditions of consistency of the bootstrap with a spectrum as proposed in 
\cite{CT} was subsequently given in \cite{TL}. Further confirmation 
from a path-integral point of view was more recently given in \cite{PRS}.

Given knowledge of conformal symmetry, spectrum and three point functions
one has in principle an unambigous construction for any genus zero
correlation function by summing over intermediate states. But
the decomposition of a n-point function as sum over three point 
functions can in general be performed in different ways. Equality of 
the expressions resulting from different such decompositions 
($\leftrightarrow$ locality, crossing symmetry) can be 
seen as being {\it the} most difficult sufficient condition to verify for
showing consistency of the CFT as characterized by spectrum
and three point functions.

The present note will outline an approach to verify 
the consistency of the bootstrap with spectrum
and three point functions as proposed in \cite{CT,DO,ZZ}. 
The main assumption underlying our approach
is the existence of duality transformations for generic 
conformal blocks that are consistent with (at least part of) the 
spectrum being continuous. It is explained that the coefficients 
that describe these transformations will then be severely constrained
by consistency conditions of Moore-Seiberg \cite{MS}\cite{FFK} type. In fact, 
taking into account known results on fusion of degenerate 
Virasoro-representations with generic ones one may derive a system of
functional equations for the fusion coefficients.
For irrational central charge $c>25$ it is possible
to show that this system of functional equations has at most one 
solution.

In order to solve these conditions, the ansatz is made that 
the fusion coefficients should be essentially given as Racah-coefficients
for an appropriate continuous series of representations of
$\USL$, with deformation parameter $q=\exp(\pi i b^2)$, 
related to the central charge $c>25$ of the Virasoro algebra via
$c=1+6(b+b^{-1})^2$. 
This is motivated by the long history of research on connections between
Liouville theory and quantum groups going back to \cite{FT,B,G} and more
recently in particular \cite{CGR,GS2,GR}.
The set of quantum group
representations considered here is unusual, however: It is neither
of lowest nor highest weight, but rather similar to the principal
continuous series of representations of $SL(2,\BR)$. Racah-coefficients
are constructed from the solution of the Clebsch-Gordan problem for
these representations, explicitly calculated and shown to provide 
a solution of all the conditions from Liouville theory discussed 
previously. In particular, the Clebsch-Gordan calculus yields orthogonality
relations for the Racah-coefficients which are just what is needed 
to establish crossing symmetry of general four point functions 
constructed in terms of the spectrum and the
three point functions proposed in \cite{CT,DO,ZZ}.

It is worth noting that the quantum group we are discussing
has no proper classical counterpart, which is related to a remarkable
self-duality under replacing the deformation parameter $q=\exp(\pi i b^2)$ by
$\tilde{q}=\exp(\pi i b^{-2})$. This symmetry directly corresponds to the 
symmetry of Liouville theory under $b\ra b^{-1}$ 
which is natural from the point of view of the bootstrap\footnote{The 
conformal blocks depend on $b$ only via $Q=b+b^{-1}$}
and encoded in the three point functions of \cite{DO, ZZ}.
\footnote{The importance of this 
self-duality for Liouville theory was also observed by L. Faddeev 
quite a while ago from a rather different point of view.}
Moreover, this duality ensures that all the relevant properties 
of the fusion coefficients will remain true when continuing to
$|b|=1$, corresponding to the strong coupling region $1<c<25$. 
The present work thus also verifies the conjecture of \cite{ZZ} that
the bootstrap remains consistent for $1<c<25$ when using the obvious
continuations of the spectrum and the three point functions that were
proposed in \cite{CT,DO,ZZ}. 

Most of our results are only announced in the present note. More details
and rigorous proofs of our quantum group results will appear in a series
of publications in preparation \cite{TQ,PT1,PT2}.

{\bf Acknowledgements:} B.P. thanks Al.B.Zamolodchikov for valuable 
discussions, and A.Neveu and G.Mennessier for taking interest in this work. 
J.T. would like to thank E. Buffenoir, L. Faddeev, V.V. Fock, P. Roche
and Al.B. Zamolodchikov for interesting discussions. 
Both authors would like to thank the organizers of the workshop
``Applications of integrability'' for the invitation and the Erwin
Schr\"{o}dinger Institute for hospitality.

This work was supported in part by the EU under 
contract ERBFMRX CT960012.

\section{Bootstrap for Liouville theory}
The possibility to preserve conformal invariance as a symmetry in the 
quantization of Liouville theory 
\cite{CT}\cite{GN} suggests to use the bootstrap formalism
\cite{BPZ} as refined in \cite{MS}\cite{FFK} in order to exploit the 
information about correlation functions that is provided by this symmetry most
efficiently. 

Conformal invariance requires that the Hilbert-space $\CH$ decomposes
as direct sum (or integral) over tensor products $\CV_{\al}\ot\CV_{\bar{\al}}$
of highest weight
representations of the left/right Virasoro algebras. The label $\al$ 
is related to the highest weight $h(\al)$ of the representation $\CV_{\al}$
via $h(\al)=\al(Q-\al)$ where $Q=b+b^{-1}$ 
is related to the central charge $c$
of the Virasoro algebra via $c=1+6Q^2$. Arguments based on canonical
quantization suggest \cite{CT} the following spectrum for Liouville theory:
\begin{equation}\label{Liouspec}
\CH=\int_{\BS}^{\oplus}d\al\;\,
\CV_{\al}\ot\CV_{\al}.\qquad \quad \BS=\frac{Q}{2}+i\BR^+
\end{equation}
Accordingly the spectrum is expected to be simple, purely continuous and 
diagonal.

The main object
of interest are correlation functions of 
Virasoro primary fields $V_{\al}(z)$, $z\in\BC$  with conformal dimension
$h(\al)$. Only the genus zero case will be considered in the
present paper. 
It should be possible to evaluate any correlation function
such as
$\bra 0| \prod_{i=1}^N V_{\al_i}(z_i)|0\ket$ by summing over intermediate
states, leading to a representation of that correlation function in terms
of the matrix elements
$\bra \al_3,d_3|V_{\al_2}(z)|\al_1,d_1\ket$. The sum over intermediate 
states splits into integrations over the intermediate 
representations and summations over vectors within fixed intermediate 
representations $\CV_{\al}\ot\CV_{\al}$. The contributions for 
fixed intermediate representations turn out to be uniquely given by 
conformal symmetry in terms of the matrix elements  
$C(Q-\al_3,\al_2,\al_1)\equiv 
\bra \al_3|V_{\al_2}(1)|\al_1\ket$ between highest weight states 
$\bra \al_3|$ and $|\al_1\ket$. In the example of the four point function
$\bra 0| \prod_{i=1}^4 V_{\al_i}(z_i)|0\ket$ one thereby arrives at a 
representation of the form
\begin{equation}\label{s-channel}\begin{aligned}
\bra 0| V_{\al_4}(z_4) & \ldots V_{\al_1}(z_1)|0\ket=\\
=& 
\int\limits_{\BS_{43|21}}
d\al_{21}\; C(\al_4,\al_3,\al_{21})C(\bal_{21},\al_2,\al_1)
\;\bigl|\CfBls{\al_{21}}{\al_3}{\al_2}{\al_4}{\al_1}(\fz)\bigr|^2,
\end{aligned}
\end{equation}
where $\bal\equiv Q-\al$ so that $\bal=\al^*$ iff $\al\in\BS$.
The {\it conformal 
blocks} $\CF_{\al_{21}}^s$ are represented by power series of the form
\begin{equation}\label{confbl2}\begin{aligned}
\CfBls{\al_{21}}{\al_3}{\al_2}{\al_4}{\al_1}(\fz)=
 & \;
z_{43}^{h_2+h_1-h_4-h_3}\;
z_{42}^{-2h_2}\;z_{41}^{h_3+h_2-h_4-h_1}
\;z_{31}^{h_4-h_1-h_2-h_3}\;\cdot\\
& \cdot z^{h(\al_{21})-h_2-h_1}
\sum_{n=0}^{\infty}z^n \CfBls{\al_{21},n}{\al_3}{\al_2}{\al_4}{\al_1},
\end{aligned}
\end{equation}
where $h_i=h(\al_i)$, $z_{ji}=z_j-z_i$, $j,i=1,\ldots,4$
and $z=\frac{z_{43}z_{21}}{z_{42}z_{31}}$. 
Conformal symmetry uniquely determines 
the coefficients  $\CF^s_{\al_{21},n}$, $n>0$ in terms of 
$\CF^s_{\al_{21},0}$ which will be chosen as unity.

\begin{rem}
One should note that the set $\BS_{43|21}$ that appears in \rf{s-channel}
will in general {\it not coincide} with the spectrum \cite{S}. 
This will only be the case when the states 
$V_{\al_2}(z_2)V_{\al_1}(z_1)|0\ket$ and $\bra 0|V_{\al_4}(z_4)V_{\al_3}(z_3)$
(suitably smeared over $z_4,\ldots,z_1$) are normalizable, which
one indeed expects \cite{S}\footnote{This can be alternatively found 
by considering asymptotics of wave-functions in some refined 
version of the canonical quantization of \cite{CT}, as will be explained 
in more detail elsewhere}
 to be the case if the parameters 
$\al_4,\ldots,\al_1$ satisfy  
\begin{equation}\label{funran}
\begin{aligned}
{}& 2|\Re(\al_1+\al_2-Q)|<Q \\
&   2|\Re(\al_3+\al_4-Q)|<Q
\end{aligned}\qquad
\begin{aligned}
{}& 2|\Re(\al_1-\al_2)|<Q \\
& 2|\Re(\al_3-\al_4)|<Q.
\end{aligned}\end{equation}
Otherwise one can have contributions from intermediate representations that
do not belong to $\BS$, but are well-defined and
uniquely determined by considering the analytic continuation of expression
\rf{s-channel} from \rf{funran} to generic complex $\al_4,\ldots,\al_1$,
as was explained in the example of the $H_3^+$-WZNW model in \cite{TH}. 
It will therefore not be a loss of generality 
to restrict attention to the 
range \rf{funran} where one indeed has $\BS_{43|21}=\BS$. 
\end{rem}

The second basic property that the operators $V_{\al}(z)$ are 
required to satisfy is mutual locality $[V_{\al}(z),V_{\be}(w)]=0$
for $z\neq w$. It follows that the four point function can 
alternatively be represented e.g. as  
\begin{equation}\label{t-channel}\begin{aligned}
\bra 0| V_{\al_4}(z_4) & \ldots V_{\al_1}(z_1)|0\ket=\\
=& 
\int\limits_{\BS}d\al_{32}\; C(\al_4,\al_{32},\al_1)C(\bal_{32},\al_3,\al_2)
\;\bigl|\CfBlt{\al_{32}}{\al_3}{\al_2}{\al_4}{\al_1}(\fz)\bigr|^2,
\end{aligned},
\end{equation}
where the {\it t-channel}\footnote{The superscript ``s'' and
``t'' refer to s- and t-channel respectively} 
conformal blocks are given by power series
similar to \rf{confbl2} with $z$ replaced by $1-z$.         

Equality of expressions \rf{s-channel} and \rf{t-channel} can be considered
as an infinite system of equations for the data $C(\al_3,\al_2,\al_1)$
and $\CS$ with coefficients $\CF^s_{\al_{21}}$, 
$\CF^t_{\al_{32}}$ given by conformal 
symmetry. As means for the practical determination of these data it 
is useless, though. 

It turns out, however, that the explicit form of the $C(\al_3,\al_2,\al_1)$
can be determined by considering certain special cases of these conditions
where the conformal blocks are known explicitly \cite{TL}. Under certain 
assumptions one finds the formula previously proposed in \cite{DO,ZZ} 
as unique solution:
\begin{equation}\label{Cexplicit}\begin{aligned}
C(\al_3 & ,\al_2 ,\al_1)= \biggl[ \pi\mu\frac{\Ga(b^2)}{\Ga(1-b^2)}b^{2-2b^2}
\biggr]^{b^{-1}(Q-\al_1-\al_2-\al_3)}\cdot\\
 & \cdot
\frac{\up_0\up_b(2\al_1)\up_b(2\al_2)\up_b(2\al_3)}{\up_b(\al_1+\al_2+\al_3-Q)
\up_b(\al_1+\al_2-\al_3)\up_b(\al_1+\al_3-\al_2)\up_b(\al_2+\al_3-\al_1)},
\end{aligned}\end{equation}
where a definition of the function $\up_b(x)$ can be found in the Appendix.

The main problem that needs to be solved in order to put the bootstrap
onto firmer ground is the verification that locality (or 
crossing symmetry) is indeed satisfied when $C(\al_3,\al_2,\al_1)$ 
as given in \rf{Cexplicit} and $\BS$ as given in \rf{Liouspec}
are used in the construction of correlation functions.
At this point we need to introduce 
our fundamental assumption: 
\begin{conj}
There exist invertible
fusion-transformations between s- and t-channel conformal blocks:
\begin{equation}\label{fusion}
\CfBls{\al_{21}}{\al_3}{\al_2}{\al_4}{\al_1}(\fz)
=\int\limits_{\BS}d\al_{32}\; 
\Fus{\al_{21}}{\al_{32}}{\al_3}{\al_2}{\al_4}{\al_1}
\;\CfBlt{\al_{32}}{\al_3}{\al_2}{\al_4}{\al_1}(\fz).
\end{equation}
\end{conj}
This conjecture is supported by an explicit calculation in a case with
a special choice of $\al_4,\ldots,\al_1$, but {\it arbitrary} $\al_{21}$ 
where an explicit expression
is available thanks to the work of A. Neveu (unpublished).

\begin{rem}
The construction of conformal blocks in terms of chiral vertex operators
\cite{MS}\cite{FFK} identifies the {\it fusion-coefficients}
$F_{\al_{21}\al_{32}}\bigl[{}^{\al_3}_{\al_4}{}^{\al_2}_{\al_1}\bigr]$ as
analogues of the Racah-Wigner or 6j-coefficients for fusion-products
of representations of the Virasoro-algebra. 
\end{rem} 

We will now restrict attention to $\al_i\in\BS$ until the end of this section.
This will not be a loss of generality since it will turn out that the
general case can be obtained by analytic continuation.
The requirement can then be rewritten as the system of equations
\begin{equation}\label{orth1}\begin{aligned}
\int\limits_{\BS}d\al_{21} \; C(\al_4,\al_3,\al_{21})C(\bal_{21},\al_2,\al_1) 
& 
\Fus{\al_{21}}{\al_{32}}{\al_3}{\al_2}{\al_4}{\al_1}
\bigl(\Fus{\al_{21}}{\be_{32}}{\al_3}{\al_2}{\al_4}{\al_1}\bigr)^*
=\\
 & =\de(\al_{32}-\be_{32})C(\al_4,\al_{32},\al_1)C(\bal_{32},\al_3,\al_2).
\end{aligned}\end{equation}
This may be brought into a more suggestive form by absorbing 
(part of) the factors
$C(\al_3,\al_2,\al_1)$ by a change of normalisation of the conformal
blocks (which is in fact a change of normalisation of the chiral 
vertex operators): Let
\begin{equation}\begin{aligned}
\CfBls{\al_{21}}{\al_3}{\al_2}{\al_4}{\al_1}(\fz)=& 
N(\al_4,\al_3,\al_{21})N(\al_{21},\al_2,\al_1)
\CgBls{\al}{\al_3}{\al_2}{\al_4}{\al_1}(\fz) \\
\CfBlt{\al}{\al_3}{\al_2}{\al_4}{\al_1}(\fz)=& 
N(\al_4,\al_{32},\al_1)N(\al_{32},\al_3,\al_2)
\CgBlt{\al_{32}}{\al_3}{\al_2}{\al_4}{\al_1}(\fz),
\end{aligned}
\end{equation}
where the following choice of $N(\al_3,\al_2,\al_1)$ will turn out to be
convenient:
\begin{equation}\label{Nexplicit}\begin{aligned}
N(\al_3 & ,\al_2,\al_1)=\\
=& \frac{\Ga_b(2\al_1)\Ga_b(2\al_2)\Ga_b(2Q-2\al_3)}
{\Ga_b(2Q-\al_1-\al_2-\al_3)
\Ga_b(\al_1+\al_2-\al_3)\Ga_b(\al_1+\al_3-\al_2)\Ga_b(\al_2+\al_3-\al_1)},
\end{aligned}\end{equation}
where $\Ga_b(x)$ is essentially the double Gamma function of Barnes \cite{Ba}, 
see the Appendix. The blocks $\CG^s_{\al}$ and $\CG^t_{\al}$ will 
then be related by an equation of the form \rf{fusion} with 
$F_{\al_{21}\al_{32}}$ replaced by
\begin{equation}\label{G-F}
\SJSL{\al_1}{\al_2}{\al_3}{\al_4}{\al_{21}}{\al_{32}}=
\frac{N(\al_4,\al_{32},\al_1)N(\al_{32},\al_3,\al_2)}{N(\al_4,\al_3,\al_{21})
N(\al_{21},\al_2,\al_1)}\;\Fus{\al_{21}}{\al_{32}}{\al_3}{\al_2}{\al_4}{\al_1}.
\end{equation}
The locality condition now takes the form
\begin{equation}\label{orth2}
\int\limits_{\BS}d\al_{21}\;\,|M_b(\al_{21})|^2 
\;\, \SJSL{\al_1}{\al_2}{\al_3}{\al_4}{\al_{21}}{\al_{32}}
\bigl(\SJSL{\al_1}{\al_2}{\al_3}{\al_4}{\al_{21}}{\be_{32}}\bigr)^*\;=\;
|M_b(\al_{32})|^2\;\,\de(\al_{32}-\be_{32}),
\end{equation}
where $M_b(\al)=-4\sin(\pi b(2\al-Q))\sin(\pi b^{-1}(2\al-Q))$.
Equation \rf{orth2} expresses {\it unitarity} of the change of 
basis \rf{fusion} when the space of conformal blocks spanned by 
$\{ \CF^s_{\al_{21}};\al_{21}\in\BS\}$ is equipped with the 
Hilbert-space structure $\bra \CF_{\al_{21}}^s,\CF_{\be_{21}}^s\ket
=|M_b(\al_{21})|^{-2}\de(\al_{21}-\be_{21})$.

It is the aim of the present work to verify \rf{orth2} by (a)
showing that the coefficients 
$G_{\al_{21}\al_{32}}\bigl[{}^{\al_3}_{\al_4}{}^{\al_2}_{\al_1}\bigr]$ 
are given 
by Racah-Wigner coefficients for a category of representations of 
$\USL$ and furthermore (b) deriving the relevant orthogonality relations
from completeness of the Clebsch-Gordan decomposition for tensor products
of these representations. 

\section{Difference equations for the fusion coefficients}
It will be assumed that the
conformal blocks for any N-point 
function can be constructed in terms of chiral vertex operators.
One therefore finds as in \cite{MS}\cite{FFK} 
that the fusion coefficients have to satisfy a system of consistency 
conditions called hexagon and pentagon equations. Here we will be 
particularly interested in the pentagon equation which takes the form
\begin{equation}\label{pentagon}
\int\limits_{\BS} d\de_1 \;\Fus{\be_1}{\de_1}{\al_3}{\al_2}{\ga_2}{\al_1}
\Fus{\be_2}{\ga_2}{\al_4}{\de_1}{\al_{5}}{\al_1}
\Fus{\de_1}{\ga_1}{\al_4}{\al_3}{\ga_2}{\al_2}
=\Fus{\be_2}{\ga_1}{\al_4}{\al_3}{\al_5}{\be_1}
\Fus{\be_1}{\ga_2}{\al_5}{\ga_1}{\al_2}{\al_1}
\end{equation}

The first crucial observation to be made at this point is that the 
fusion coefficients should have certain analyticity
properties in the dependence on its six complex
parameters. Note that the coefficients 
$\CF_{\al_{21},n}^{s}$ that appear in the 
power series representation of conformal blocks \rf{confbl2} depend
{\it polynomially} on the variables $\al_4,\ldots,\al_1$ and rationally
on $\al$, with poles only at the values 
$2\al=\al_{m,n}=-mb-nb^{-1}$ and $Q-\al=\al_{m,n}$. 
If the
series \rf{confbl2} converges\footnote{Liouville theory would be dead
alltogether otherwise} it follows that
the dependence of $\CF_{\al_{21}}^{s}$ on the
variables $\al_4,\ldots,\al_1$ must be entire analytic, whereas the
$\al$-dependence will be meromorphic with poles only at the locations
given previously. The compatibility of the existence of
fusion transformations \rf{fusion}, analyticity properties
of conformal blocks
and Remark 1 suggests (see \cite{PT2} for more discussion) the following 
conjecture:
\begin{conj}
The fusion coefficients 
$F_{\al_{21}\al_{32}}\bigl[{}^{\al_3}_{\al_4}{}^{\al_2}_{\al_1}\bigr]$
are holomorphic in 
\begin{equation}\label{funran2}
\begin{aligned}
0<&\Re(\al_2+\al_3+\al_{23}-Q)<Q\\
0<&\Re(\al_2+\al_3-\al_{23})<Q\\
0<&\Re(\al_{23}+\al_3-\al_2)<Q\\
0<&\Re(\al_{23}+\al_2-\al_3)<Q
\end{aligned}\qquad
\begin{aligned}
0<&\Re(\al_4+\al_1+\al_{23}-Q)<Q\\
0<&\Re(\al_4+\al_1-\al_{23})<Q\\
0<&\Re(\al_{23}+\al_4-\al_1)<Q\\
0<&\Re(\al_{23}+\al_1-\al_4)<Q
\end{aligned}
\end{equation}
\end{conj}
Second, one may observe that the fusion transformations \rf{fusion}
simplify if one of $\al_4,\ldots,\al_1$ is taken to be 
$\al=\al_{m,n}=-mb-nb^{-1}$, corresponding to a degenerate
representation of the Virasoro algebra. Conformal
blocks $\CF^s_{\al_{21}}$, $\CF^t_{\al_{32}}$ 
then only exist for a finite number 
of values of $\al_{21}$, $\al_{32}$, 
so that the fusion coefficients form a finite
dimensional matrix \cite{BPZ}. 
It will suffice to consider cases where one of 
$\al_4,\ldots,\al_1$, say $\al_2$, equals $-b$ or $-b^{-1}$. In that
case s-channel conformal blocks $\CF^s_{\al_{21}}$ exist only for 
$\al=\al_1-sb$, $s=-,0,+$, and t-channel conformal blocks for
$\al=\al_3-sb$, $s=-,0,+$. Moreover, it is easy to show that
the fusion coefficients that appear in such cases can all be uniquely
expressed in terms of the following ``elementary'' ones:
Let 
\begin{equation} F_{s,s'}(\al_4,\al_3,\al_1)\equiv
 \Fus{\al_1-s\frac{b}{2}}{,\al_3-s'\frac{b}{2}}{\al_3}{\al_2}
{\al_4}{\al_1}_{\al_2=-\frac{b}{2}}, \qquad\text{where}\quad s,s'=+,-
\end{equation}
The matrix $F_{s,s'}$ can be calculated explicitly \cite{GN}\cite{BPZ}.
The corresponding matrix $G_{s,s'}$ (cf. \rf{G-F}) is then given by
\begin{equation}
\begin{aligned}
G_{++}=& \frac{[\al_4+\al_3-\al_1-\frac{b}{2}]}
{[2\al_3-b]}\\
G_{-+}=& \frac{[\al_3+\al_1-\al_4-\frac{b}{2}]}
{[2\al_3-b]}
\end{aligned}\qquad
\begin{aligned}
G_{+-}=& \frac{[\al_4+\al_3+\al_1-\frac{3b}{2}]}
{[2\al_3-b]}\\
G_{--}=& -\frac{[\al_4+\al_1-\al_3-\frac{b}{2}]}
{[2\al_3-b]}
\end{aligned}\qquad [x]\equiv \frac{\sin(\pi b x)}{\sin(\pi b^2)}.
\end{equation}

 If one then considers the pentagon equation \rf{pentagon} in the special cases
where one of $\al_5,\ldots,\al_1$ equals $-b$, one finds a set of 
{\it linear finite difference} equations for the general fusion coefficients
$F_{\al_{21}\al_{32}}\bigl[{}^{\al_3}_{\al_4}{}^{\al_2}_{\al_1}\bigr]$. 
Part of these equations involve shifts of one argument only,
for example
\begin{equation}\label{homog}
\sum_{s=-,0,+}C_{s}\displaystyle{\bigl( } \ew 
\begin{array}{ll} {\scriptstyle \al_1}
  \ppe & {\scriptstyle \al_2} \ppe \\[-2mm] {\scriptstyle \al_3}\ppe &
  {\scriptstyle \al_4}\ew \end{array}\big| \ew
\begin{array}{l} {\scriptstyle \al_{21} }
  \ppe \\[-2mm] {\scriptstyle \al_{32}}\ew  \end{array}\displaystyle{\bigr)}
\Fus{\al_{21}}{\al_{32}}{\al_3}{\al_2}
{\al_4}{\al_1-sb}=0.
\end{equation}
One has one such equations of each of the variables $\al_4,\ldots\al_1$. 
Other equations are of the form
\begin{equation}\label{inhomog}
\sum_{s=-,0,+}D_{r,s}\displaystyle{\bigl( } \ew 
\begin{array}{ll} {\scriptstyle \al_1}
  \ppe & {\scriptstyle \al_2} \ppe \\[-2mm] {\scriptstyle \al_3}\ppe &
  {\scriptstyle \al_4}\ew \end{array}\big| \ew
\begin{array}{l} {\scriptstyle \al_{21} }
  \ppe \\[-2mm] {\scriptstyle \al_{32}}\ew  \end{array}\displaystyle{\bigr)}
\Fus{\al_{21}}{\al_{32}}{\al_3}{\al_2}
{\al_4}{\al_1-sb}=\Fus{\al_{21}+rb}{,\al_{32}}{\al_3}{\al_2}
{\al_4}{\al_1-sb}\quad\text{where $r=+,-$},
\end{equation}
and a similar equation with shifts of $\al_{32}$ on the right hand side.
Furthermore, each of these equations has a ``dual'' partner obtained by 
$b\ra b^{-1}$. Finally, one has equations that reflect the fact that 
all the fusion coefficients $F_{\be_1,\be_2}
\bigl[{}^{\al_3}_{\al_4}{}^{\al_2}_{\al_1}\bigr]$ are functions 
of the conformal dimensions only, so must be unchanged under 
$\al_i\ra Q-\al_i$, $i=1,2,3,4$ and $\be_j\ra Q-\be_j$, $j=1,2$.

In the case of real irrational $b$ it is possible to show 
(details will appear in \cite{PT2}) uniqueness
of a solution to this system of functional equations, 
taking into account the analytic properties 
of the fusion coefficients. In fact, the equations \rf{homog} are second
order homogeneous finite difference equations. 
It can be shown  
that the second order equations of the form \rf{homog}
together with their $b\ra b^{-1}$ duals can have at most two linearly
independent solutions with the required analytic properties. Taking 
into account the symmetry $\al_i\ra Q-\al_i$, $i=1,\ldots,4$
will determine the dependence w.r.t. $\al_i$, $i=1,\ldots,4$ up to 
a factor that depends on $\al_{21}$ and $\al_{32}$. The remaining
freedom is then fixed by considering equations \rf{inhomog} and its 
counterpart with shifts of $\al_{32}$.

\begin{rem}
Loosely speaking the message is the following:
If there exist fusion transformations of conformal blocks
that are compatible with the expectations from other approaches
(encoded in Conjectures 1 and 2; cf. Remark 1) 
then they are unique for real, irrational $b$.
\end{rem}

\begin{rem}
One might also be interested in the possibility of having fusion
transformations of the form \rf{fusion} but with coefficients 
$F_{\al_{21},\al_{32}}
\bigl[{}^{\al_3}_{\al_4}{}^{\al_2}_{\al_1}\bigr]$
only required to be defined for real $\al_4,\ldots,\al_1$
{\it and} $\al_{21},\al_{32}$. But if one then only requires e.g. continuity
in some interval such as \rf{funran2} one still has the above result
on uniqueness, which together with the results to be discussed below
put one back precisely into the situation considered here. 
\end{rem}

\section{A tensor category of quantum group representations}
A set of infinite dimensional representations $\CP_{\al}$ of the quantized 
universal enveloping algebra $\CU_q(\fsl(2,\BR))$ may be realized on the
Hilbert space $L^2(\BR)$ in terms of the Weyl-algebra generated
by $U=e^{2\pi b x}$ and $V=e^{-\frac{b}{2}p}$, where $[x,p]=i$:
\begin{equation}\label{uslgens}
\begin{aligned}
E=U^{+1}\frac{e^{\pi i b (Q-\al)}V- e^{-\pi i b (Q-\al)}V^{-1}}{e^{\pi i b^2}-
e^{-\pi i b^2}}\\
F=U^{-1}\frac{e^{-\pi i b (Q-\al)}V- e^{\pi i b (Q-\al)}V^{-1}}{e^{\pi i b^2}-
e^{-\pi i b^2}}
\end{aligned}\qquad\qquad K=V
\end{equation}
These generators satisfy the relations 
\begin{equation}
KE=qEK\qquad\quad KF=q^{-1}FK\qquad\quad [E,F]=-\frac{K^2-K^{-2}}{q-q^{-1}}
\end{equation}
The operators $E$, $F$, $K$ are unbounded. They will be defined on domains
consisting of functions which possess an analytic continuation into the
strip $\{x\in\BC;|\Im(x)|<\frac{b}{2}\}$ and which have suitable
decay properties at infinity.
It may be shown that the representations $\CP_{\al}$, $\al\in\BS$
are unitarily equivalent
to a subset of the integrable (``well-behaved'') representations
of $\CU_q(\fsl(2,\BR))$ classified in \cite{S2}. It follows in particular
that $E$, $F$ and $K$ become self-adjoint on suitable domains.

The q-Casimir acts as a scalar in this representation:
\begin{equation}
C=FE-\frac{qK^2+q^{-1}K^{-2}-2}{(q-q^{-1})^2}\equiv [\al-\fr{Q}{2}]^2,
\qquad [x]\equiv \frac{\sin(\pi b x)}{\sin(\pi b^2)}.
\end{equation}

\begin{rem}
The representations considered here 
form a subset of the representations
of $\CU_q(\fsl(2,\BR))$ that appear in the classification of \cite{S2}. 
This subset is distinguished by the fact that it is simultaneously
a representation of $\CU_{\tilde{q}}(\fsl(2,\BR))$, 
$\tilde{q}=\exp(\pi i b^{-2})$
with generators $\tE$, $\tF$, $\tK$ being realized by replacing $b\ra b^{-1}$
in the
expressions for $E$, $F$, $K$ given above. Restriction
to these representations is crucial for obtaining a quantum group
structure which
is self-dual under $b\ra b^{-1}$, as Liouville theory is. The price to pay
is that the representations $\CP_{\al}$ do not have classical ($b\ra 0$) 
counterparts.
\end{rem}

Tensor products $\CP_{\al_2}\ot\CP_{\al_1}$ of representations 
can be defined by means of the co-product:
\begin{equation}
\De(K)=K\ot K\qquad \De(E)=E\ot K^{-1}+K\ot E\qquad 
\De(F)=F\ot K^{-1}+K\ot F
\end{equation}

\begin{thm}\label{Cldeco}
The $\CU_{q}(\fsl(2,\BR))$-representation $\CP_{\al_2}\ot\CP_{\al_1}$ 
defined on $L^2(\BR^2)$ by means of $\De$ decomposes as follows
\begin{equation}\label{CGdecoQG}
\CP_{\al_2}\ot \CP_{\al_1}\simeq \int_{\BS}^{\oplus}
\!d\al\;\, \CP_{\al,}
\end{equation}\end{thm}
It is remarkable and nontrivial that the subset of ``self-dual'' integrable
representations of $\USL$ is actually closed under tensor products. 
The proof, based on results and 
techniques of \cite{TQ}, will appear in \cite{PT1}.

The Clebsch-Gordan maps $C(\al_3|\al_2,\al_1):\CP_{\al_2}\ot\CP_{\al_1}\ra
\CP_{\al_3}$ may be explicitly represented as an integral transform
\begin{equation}
C(\al_3|\al_2,\al_1): f(x_2,x_1)\longrightarrow F[f](\al_3|x_3)
\equiv \int_{\BR}dx_2dx_1 \;\CGC{\al_3}{x_3}{\al_2}{x_2}{\al_1}{x_1}\;
f(x_2,x_1).
\end{equation}
The distributional kernel $[\ldots]$ (the "Clebsch-Gordan coefficients")
is given by the expression
\begin{equation}\label{Clebsch}\begin{aligned}
\CGC{Q-\al_3}{x_3}{\al_2}{x_2}{\al_1}{x_1}
= & S_b(\al_1+\al_2-\al_3)e^{\pi i \al_1\al_2}\;\,
e^{2\pi (x_3(\al_2-\al_1)-\al_2x_2+\al_1x_1)}\cdot \\
& \cdot 
D_b(x_{32},\al_{32})D_b(x_{31},\al_{31})D_b(x_{21},\al_{21}) 
\end{aligned}\end{equation}
where the distribution $D_b(x,\al)$ is defined in terms of the 
Double Sine function $S_b(x)$ (cf. Appendix) as
\begin{equation}
D_b(x,\al)=e^{-\frac{\pi i}{2}a(a-Q)}
e^{\pi a x}\lim_{\ep\ra0+}\frac{S_b(ix+\ep)}{S_b(ix+\al)}
\end{equation}
and the coefficients $x_{ji}$, $\al_{ji}$, $j>i\in\{1,2,3\}$ are given by
\begin{equation}\begin{aligned}
x_{32}=& x_3-x_2+\fr{i}{2}(\al_3+\al_2-Q)\\
x_{31}=& x_3-x_1+\fr{i}{2}(\al_3+\al_1-Q)\\
x_{21}=& x_3-x_2+\fr{i}{2}(\al_2+\al_1-2\al_3)
\end{aligned}
\qquad \quad \begin{aligned}
\al_{32}=& \al_2+\al_3-\al_1\\
\al_{31}=& \al_3+\al_1-\al_2\\
\al_{21}=& \al_2+\al_3-\al_1.
\end{aligned}
\end{equation}
The orthogonality relations for the Clebsch-Gordan coefficients \rf{Clebsch}
can be determined by explicit calculation: 
\begin{equation}\label{CGorth}
\int\limits_{\BR}dx_1dx_2\;\,
\CGC{\al_3}{x_3}{\al_2}{x_2}{\al_1}{x_1}^*
\CGC{\be_3}{y_3}{\al_2}{x_2}{\al_1}{x_1}
=|S_b(2\al_3)|^{-2}\de(\al_3-\be_3)\de(x_3-y_3).\end{equation}
Together with Theorem \ref{Cldeco} one obtains the corresponding completeness
relations
\begin{equation}\label{CGcompl}
 \int\limits_{\BS}d\al_3 \; |S_b(2\al_3)|^2
\int\limits_{\BR}dx_3 \;\,
\CGC{\al_3}{x_3}{\al_2}{x_2}{\al_1}{x_1}^*
\CGC{\al_3}{x_3}{\al_2}{y_2}{\al_1}{y_1}
=\de(x_2-y_2)\de(x_1-y_1). 
\end{equation}

The braiding-operation $B:\CP_{\al_2}\ot\CP_{\al_1}\ra 
\CP_{\al_1}\ot\CP_{\al_2}$ may then be introduced as
\footnote{L. Faddeev has explained to the authors a nice alternative
method to introduce an R-operator.}
\begin{equation}
B_{21}\;=\;
\int\limits_{\BS}d\al_3 \;\, |S_b(2\al_3)|^2 \;\,C^{\dagger}(\al_1,\al_2|\al_3)
\;\OM{\al_3}{\al_2}{\al_1}\;C(\al_3|\al_2,\al_1),
\end{equation}
where $C^{\dagger}(\al_1,\al_2|\al_3): \CP_{\al_3}\ra \CS_{21}'$
is the adjoint of $C(\al_3|\al_2,\al_1)$ for any Gelfand-triple 
$\CS_{21}\subset\CP_{\al_2}\ot\CP_{\al_1}\subset\CS_{21}'$,
and 
\begin{equation}
\OM{\al_3}{\al_2}{\al_1}=
e^{\pi i(\al_3(Q-\al_3)-\al_2(Q-\al_2)-\al_1(Q-\al_1))}.
\end{equation}

Triple tensor products $\CP_{\al_3}\ot
\CP_{\al_2}\ot\CP_{\al_1}$ carry a representation of $\USL$ given by 
$(\id\ot\De)\circ\De=(\De\ot\id)\circ\De$.
The projections affecting the decomposition of this representation into
irreducibles can be constructed by iterating Clebsch-Gordan maps.
One thereby obtains two canonical bases
in the sense of generalized eigenfunctions for $\CP_{\al_3}\ot
\CP_{\al_2}\ot\CP_{\al_1}$ given by the sets of distributions 
($\fx=(x_4,\ldots,x_1)$)
\begin{equation}\label{q-blocks}
\begin{aligned}
\CpBls{\al_{21}}{\al_3}{\al_2}{\al_4}{\al_1}(\fx)=& 
\int_{\BR}dx_{21}\;\,
\CGC{\al_4}{x_4}{\al_3}{x_3}{\al_{21}}{x_{21}}\,
\CGC{\al_{21}}{x_{21}}{\al_2}{x_2}{\al_1}{x_1}
\quad\al_4,\al_{21}\in\BS,x_4\in\BR\\
\CpBlt{\al_{32}}{\al_3}{\al_2}{\al_4}{\al_1}(\fx)=& 
\int_{\BR}dx_{32}\;\,
\CGC{\al_4}{x_4}{\al_{32}}{x_{32}}{\al_1}{x_1}\,
\CGC{\al_{32}}{x_{32}}{\al_3}{x_3}{\al_2}{x_2}.
\quad\al_4,\al_{32}\in\BS,x_4\in\BR\end{aligned}.
\end{equation}
In particular one may observe that $\CP_{\al_3}\ot
\CP_{\al_2}\ot\CP_{\al_1}$
can be decomposed into eigenspaces of 
the pair of commuting self-adjoint operators
$\pi_{321}(C)$ and $\pi_{321}(K)$ as follows:
\begin{equation}
\CP_{a_3}\ot\CP_{a_2}\ot \CP_{a_1}\simeq 
\int\limits_{\BS}^{\oplus}d\al\int\limits_{\BR}^{\oplus}dk
\;\, \CH_{\al,k}.
\end{equation}
It then follows from completeness of the bases $\FB_{321}^s$ and $\FB_{321}^t$
and orthogonality of the eigenspaces $ \CH_{\al,k}$ that 
the bases $\Phi^s$ and $\Phi^t$ must be related by a transformation of the
form
\begin{equation}\label{Racahdef}
\CpBls{\al_{21}}{\al_3}{\al_2}{\al_4}{\al_1}(\fx)\;=\;
\int\limits_{\BS}d\al_{32}\;\,
\SJS{\al_1}{\al_2}{\al_3}{\al_4}{\al_{21}}{\al_{32}}\;\,
\CpBlt{\al_{32}}{\al_3}{\al_2}{\al_4}{\al_1}(\fx)
\end{equation}
thereby defining the b-Racah-Wigner symbols $\SJS{\cdot}{\cdot}
{\cdot}{\cdot}{\cdot}{\cdot}$.

The data $\OM{\al_3}{\al_2}{\al_1}$ and 
$\bigl\{ {}^{\al_1}_{\al_3}{}^{\al_2}_{\al_3}| 
{}^{\al_{21}}_{\al_{32}}\bigr\}_b$
will now satisfy all the Moore-Seiberg consistency conditions.

Moreover, by again 
using completeness of the bases $\FB_{321}^s$ and $\FB_{321}^t$
one finds the following orthogonality relations for the b-Racah-Wigner symbols
\begin{equation}\label{orth3}
\int\limits_{\BS}d\al_{21}\;|S_b(2\al_{21})|^2 
\; \SJS{\al_1}{\al_2}{\al_3}{\al_4}{\al_{21}}{\al_{32}}
\bigl(\SJS{\al_1}{\al_2}{\al_3}{\al_4}{\al_{21}}{\be_{32}}\bigr)^*\;=\;
|S_b(2\al_{32})|^2\;\de(\al_{32}-\be_{32}).
\end{equation}
In fact, $|S_b(2\al)|^2=|M_b(\al)|^2$, so that \rf{orth3} is indeed 
the orthogonality relation required to prove locality or crossing 
symmetry in Liouville theory once equality of 
$G_{\al_{21},\al_{32}}\bigl[{}^{\al_3}_{\al_4}{}^{\al_2}_{\al_1}\bigr]$ and 
$\bigl\{ {}^{\al_1}_{\al_3}{}^{\al_2}_{\al_3}| 
{}^{\al_{21}}_{\al_{32}}\bigr\}_b$
will be established.

\section{Space of functions on the quantum group}

The following section may be skipped by readers interested mainly in Liouville
theory. It is important, however, for the deeper mathematical
understanding why there is such a remarkable non-classical
category of representations. An explanation can be given by constructing
the associated dual object (a suitable space of functions on the
corresponding quantum group) and studying its harmonic analysis.
Details will be given in \cite{TQ}.

Define operators $A,B,C,D$ on $L^2(\BR\ti\BR)$ by the expressions
\begin{equation}
\begin{aligned}
A= & U_1^{-1} \\
C= & V_1 
\end{aligned}
\qquad 
\begin{aligned}
B= & V_1U_2 \\
D= & U_1(1+qU_2V_1^2). 
\end{aligned}\qquad \begin{aligned}
U_j=& e^{2\pi b x_j} \\
V_j=& e^{-\frac{b}{2}p_j}\end{aligned}
\qquad [x_i,p_j]=i\de_{i,j}\qquad j=1,2
\end{equation}
This is known to generate an integrable operator representation of the 
algebra
$\text{Pol}(SL_q(2,\BR))$ \cite{S2}. In particular, the operators $A,B,C,D$
all have self-adjoint extensions and one has the nontrivial relations
\begin{equation}\label{Polrels}\begin{aligned}
AB= & qBA \\
AC= & qCA \end{aligned}
\qquad 
\begin{aligned}
DB= & q^{-1}BD \\
DC= & q^{-1}CD \end{aligned}\qquad\begin{aligned}
AD-&qBC= 1\\
AD-&DA=(q-q^{-1})BC.
\end{aligned}\end{equation}
Again it is worth noting that restriction to the integrable representations
of $\text{Pol}(SL_q(2,\BR))$ from \cite{S2} in which $A,B,C,D$ have 
positive spectrum
already represents a point of departure from classical $SL(2,\BR)$ 
\footnote{The spectrum does not cover the group manifold.}, but has the
desired feature of being simultaneously a representation of the
($b\ra b^{-1}$)-dual algebra $\text{Pol}(SL_{\tilde{q}}(2,\BR))$. 
The algebra generated by {\it positive selfadjoint} Hilbert space operators 
$A,B,C,D$ with relations \rf{Polrels} will be denoted 
$\text{Pol}(SL_{q}^+(2,\BR))$.

Let $\CA$ be the norm closure of the set $\CC_{c}^{\infty}(SL_q^+(2,\BR))$
of all operators on $L^2(\BR\ti\BR)$
of the form
\begin{equation}\label{PDO}
\CO[f]=\int\limits_{\BR}drds \;\,A^{\frac{ir}{b}} 
\;C^{\frac{is}{b}}f(r,s|x)\;C^{\frac{is}{b}}\; A^{\frac{ir}{b}},\qquad
 x\equiv \fr{1}{2\pi b}\log(BC)
\end{equation}
where the so-called {\it symbol} 
$f(r,s|x)$ is smooth and of compact support in its dependence of
the variable $x$, 
and of Paley-Wiener class (entire analytic, rapid decay) w.r.t. 
the variables $r$ and $s$. $\CA$ is then a nonunital $C^*$-algebra 
generated by elements $A,B,C,D$ affiliated with $\CA$ in the sense of 
Woronowicz \cite{W1,W2}.

On the Hopf-algebra $\text{Pol}(SL_q(2,\BR))$ one has natural analogues of
left- and right regular representation defined by means of the duality
between $\text{Pol}(SL_q(2,\BR))$ and $\USL$, see e.g. \cite{MMNNSU}.
There is a corresponding action on $\CC_{c}^{\infty}(SL_q^+(2,\BR))$ that
may be expressed conveniently in terms of finite difference operators
acting on the symbols $f(r,s|x)$:
\begin{equation}
\begin{aligned}
E_l=& T_r^+T_s^+e^{-2\pi b x}[\de_x] \\
F_l=& T_r^-T_s^-([\de_x+2is]+e^{2\pi b x}[\de_x+2i(s+r)])\\
E_r=& T_r^-T_s^+([\de_x+2ir]+e^{-2\pi b x}[\de_x]) \\
F_r=& T_r^+T_s^-[\de_x+2is]
\end{aligned}\qquad\begin{aligned}
 K_l=& e^{-\pi(r+s)}\\ K_r=& e^{-\pi(r-s)},\end{aligned}
\end{equation}
where the following notation has been used: 
$[x]\equiv \frac{\sin(\pi b x)}{\sin(\pi b^2)}$ and 
\begin{equation}
T_r^{\pm}f(r,s|x)=f(r\pm\fr{ib}{2},s|x)\qquad T_s^{\pm}f(r,s|x)=f(r,s\pm\fr{ib}{2}|x)\qquad \de_x=\fr{1}{2\pi}\pa_x.
\end{equation}

A $L^2$-space may be introduced as completion of $\CC_{c}^{\infty}(SL_q(2,\BR))$ with respect to the inner product $\bra.,.\ket$ defined as
\begin{equation}\label{Haar-measure}
\bra \CO[f_1],\CO[f_2]\ket =\int\limits_{\BR}dr ds \int\limits_{\BR}dx\; 
e^{2\pi Qx} \bigl(f_1(r,s|x)\bigr)^*f_2(r,s|x).
\end{equation}
This inner product is such that the operators $E_l,F_l,K_l$
and $E_r,F_r,K_r$ are symmetric. 

\begin{thm} One has the following decomposition of 
$L^2(SL_q^+(2,\BR))$ into irreducible representations of 
$\USL_l\ot\USL_r$: 
\begin{equation}\label{Plancherel}
L^2(SL_q^+(2,\BR))\simeq \int\limits_{\BS}^{\oplus}
\!d\al\;|S_b(2\al)|^2 \;\,\CP_{\al}\ot\CP_{\al}
\end{equation}
\end{thm}

One of the authors (J.T.) has partial results that strongly support the
conjecture that the co-product exists on the Hilbert-space level in the 
sense of \cite{W1}.

\begin{rem}One obtains representations of $\USL_l\ot\USL_r$ by symmetric
operators $E_l,F_l,K_l$
and $E_r,F_r,K_r$ also if one chooses the measure in \rf{Haar-measure} to be 
$\exp(2\pi (b+\frac{k}{b}) x)$. The choice $k=0$ in particular reproduces
the Haar-measure on classical $SL(2,\BR)$. For the present choice $k=1$ 
one looses the correspondence to any classical object but gains the
self-duality $b\ra b^{-1}$. 
\end{rem}
\section{Calculation of Racah coefficients}
The direct calculation of the Racah coefficients from the Clebsch-Gordan
maps looks difficult. A small trick helps. If one fixes the values of 
three of the four variables $x_4,\ldots,x_1$ in \rf{Racahdef} one obtains
an integral transformation for a function of a single variable. In fact, the
analytic properties of $\Phi_{\al_{21}}^s$ and $\Phi_{\al_{32}}^t$ even allow
to choose complex values. It will be convenient to consider
\begin{equation} \label{limits}
\CBls{\al_{21}}{\al_3}{\al_2}{\bar{\al}_4}{\al_1}{\Psi}(x)=
\lim_{x_4\ra\infty}e^{2\pi\al_4x_4}
\lim_{x_2\ra-\infty}
\prod_{j=1}^{3}e^{-2\pi \al_jx_j}
\CBls{\al_{21}}{\al_3}{\al_2}{\bar{\al}_4}{\al_1}{\Phi}(\fx)
\Bigr|_{x_3=\frac{i}{2}(Q+\al_2-\al_4)}^{x_1=x},
\end{equation} 
and the same for $\Psi^t_{\al_{32}}$. The integral that defines 
$\Phi_{\al_{21}}^s$ and $\Phi_{\al_{32}}^t$ according to 
\rf{q-blocks},\rf{limits} can be done explicitly by using \rf{Eulerint}. 
One finds expressions of the form
\begin{equation}\begin{aligned}
{} & \CBls{\al_{21}}{\al_3}{\al_2}{\bar{\al}_4}{\al_1}{\Psi}(x)= 
\CBls{\al_{21}}{\al_3}{\al_2}{\bar{\al}_4}{\al_1}{N}
\CBls{\al_{21}}{\al_3}{\al_2}{\bar{\al}_4}{\al_1}{\Theta}(x)\\
& \qquad \CBls{\al_{21}}{\al_3}{\al_2}{\bar{\al}_4}{\al_1}{\Theta}(x)=
e^{+2\pi x(\al_{21}-\al_2-\al_1)}
F_b(\al_{21}+\al_1-\al_2,\al_{21}+\al_3-\al_4;2\al_{21};-ix)\\
 & \CBlt{\al_{32}}{\al_3}{\al_2}{\bar{\al}_4}{\al_1}{\Psi}(x)= 
\CBlt{\al_{32}}{\al_3}{\al_2}{\bar{\al}_4}{\al_1}{N}
\CBlt{\al_{32}}{\al_3}{\al_2}{\bar{\al}_4}{\al_1}{\Theta}(x)\\
& \qquad\CBlt{\al_{32}}{\al_3}{\al_2}{\bar{\al}_4}{\al_1}{\Theta}(x)=
e^{-2\pi x(\al_{32}+\al_1-\al_4)}
F_b(\al_{32}+\al_3-\al_2,\al_{32}+\al_1-\al_4;2\al_{32};+ix),
\end{aligned}
\end{equation}
where $F_b$ is the b-hypergeometric function defined in the Appendix,
and $N_{\al_{21}}^s$, $N_{\al_{32}}^t$ are certain normalization factors.

The linear transformation following
from \rf{Racahdef} can now be calculated as follows: One observes that
$\Psi^s_{\al_{21}}$ (resp. $\Psi^t_{\al_{32}}$) are eigenfunctions
of the finite difference operators $\CC_{21}$ and $\CC_{32}$
defined respectively by
\begin{equation}\begin{aligned}
\CC_{21}= & \bigl[\de_x+\al_1+\al_2-\fr{Q}{2}\bigr]^2-
e^{+2\pi b x}\bigl[\de_x+\al_1+\al_2+\al_3-\al_4\bigr]
\bigl[\de_x+2\al_1\bigr]\\
 \CC_{32}= & \bigl[\de_x+\al_1-\al_4+\fr{Q}{2}\bigr]^2-
e^{-2\pi b x}\bigl[\de_x+\al_1+\al_2-\al_3-\al_4\bigr]\bigl[\de_x\bigr],
\end{aligned}\end{equation}
where $\de_x=(2\pi)^{-1}\pa_x$. These operators can be made self-adjoint
in $L^2(\BR,dx e^{2\pi Qx})$, and it can be shown that 
\begin{thm}
$\{ \Theta^s_{\al_{21}};\al_{21}\in\BS\}$ and 
$\{ \Theta^t_{\al_{32}};\al_{32}\in\BS\}$ form complete 
sets of eigenfunctions of the operators $\CC_{21}$ and $\CC_{32}$ 
respectively, normalized by 
\begin{equation}
\int\limits_{\BR} dx \;e^{2\pi Qx}\; \bigl(
\CBls{\al_{21}'}{\al_3}{\al_2}{\bar{\al}_4}{\al_1}{\Theta}(x) \bigr)^* 
\CBls{\al_{21}}{\al_3}{\al_2}{\bar{\al}_4}{\al_1}{\Theta}(x)=
\de(\al_{21}-\al_{21}').
\end{equation}\end{thm}
It follows that the Racah-Wigner coefficients can be evaluated
in terms of the overlap between these two bases:
\begin{equation}
\SJS{\al_1}{\al_2}{\al_3}{\bal_4}{\al_{21}}{\al_{32}}\;=\;
\frac{\CBls{\al_{21}}{\al_3}{\al_2}{\bar{\al}_4}{\al_1}{N}}
{\CBlt{\al_{32}}{\al_3}{\al_2}{\bar{\al}_4}{\al_1}{N}}
\;\,\int\limits_{\BR} dx \;e^{2\pi Qx}\; \bigl(
\CBlt{\al_{32}}{\al_3}{\al_2}{\bar{\al}_4}{\al_1}{\Theta}(x) \bigr)^* 
\CBls{\al_{21}}{\al_3}{\al_2}{\bar{\al}_4}{\al_1}{\Theta}(x).
\end{equation}
The integral can be done by using the representation \rf{Barnesint}
for the b-hypergeometric function. The result is 
\begin{equation} \begin{aligned}
{} & \SJS{\al_1}{\al_2}{\al_3}{\bar{\al}_4}{\al_{21}}{\al_{32}} =\\
& \quad\frac{S_b(\al_2+\al_{21}-\al_1)S_b(\al_2+\al_1-\al_{21})
S_b(\al_{21}+\al_3+\al_4-Q)S_b(\al_{32}+\al_1-\al_4)}
     {S_b(\al_2+\al_{32}-\al_3)S_b(\al_3+\al_2-\al_{32})
S_b(\al_{21}+\al_3-\al_4)S_b(\al_{32}+\al_1+\al_4-Q)}\cdot\\
&\qquad\cdot|S_b(2\al_{32})|^2
\int\limits_{-i\infty}^{i\infty}ds \;\;
\frac{S_b(U_1+s)S_b(U_2+s)S_b(U_3+s)S_b(U_4+s)}
{S_b(V_1+s)S_b(V_2+s)S_b(V_3+s)S_b(V_4+s)},
\end{aligned} 
\end{equation}
where the coefficients $U_i$ and $V_i$, $i=1,\ldots,4$ are given by
\begin{equation}  
\begin{aligned} U_1=& \al_{21}+\al_1-\al_2 \\
        U_2=& Q+\al_{21}-\al_2-\al_1 \\
        U_3=& \al_{21}+\al_3-\al_4 \\
        U_4=& Q+\al_{21}-\al_3-\al_4
\end{aligned}\qquad
\begin{aligned} 
        V_1=& 2Q+\al_{21}-\al_{32}-\al_2-\al_4 \\
        V_2=& Q+\al_{21}+\al_{32}-\al_4-\al_2 \\
        V_3=& 2\al_{21} \\
        V_4=& Q.
\end{aligned}
\end{equation}
The integral representing the Racah-Wigner
coefficients is of the type of a Barnes integral
for a ${}_4F_3$ b-hypergeometric function.

\section{Functional equations for Racah-Wigner coefficients}
There is a family of finite-dimensional representations 
of $\USL\ot \CU_{\tilde{q}}(\fsl(2,\BR))$ 
labelled by two positive integers $n,m$. The generators
$E,F,K$ and $\tE,\tF,\tK$ are realized on vector spaces $\CP_{n,m}$
of polynomials 
in the variables $U$ and $\tU=e^{2\pi b^{-1}x}$ by means of
the restriction of the expressions \rf{uslgens} to 
$Q-\al=-\frac{n}{2}b-\frac{m}{2}b^{-1}$, $n,m=0,1,2,\ldots$.
Of particular interest will be the pair of two-dimensional representations
$\CP_{1,0}$ and $\CP_{0,1}$, from which all representations $\CP_{n,m}$ can
be generated by repeated tensor products. The decomposition of tensor 
products of representations $\CP_{1,0}$ or $\CP_{0,1}$ with a generic 
representation $\CP_{\al}$ into irreducible
representations can be determined purely algebraically:
\begin{equation}
\CP_{1,0}\ot\CP_{\al}\simeq\bigoplus_{s=\pm}\CP_{\al-s\frac{b}{2}}\qquad\quad
\CP_{0,1}\ot\CP_{\al}\simeq\bigoplus_{s=\pm}\CP_{\al-s\frac{1}{2b}}
\end{equation}

The distributions $\Psi^s_{\al_{21}}$ (resp. $\Psi^t_{\al_{32}}$) develop
double poles if one sets e.g.
$\al_2=-\frac{b}{2}$, $\al_{21}=\al_1-\si\frac{b}{2}$, $\si=+,-$
(resp. $\al_{32}=\al_3-\si\frac{b}{2}$ and $b\ra b^{-1}$). The 
relevant coefficients describing tensoring with the finite dimensional
representations $\CP_{1,0}$ (resp. $\CP_{0,1}$) are found as residues
of these double poles. These residues will be denoted as 
$\Psi^s_{\si}$ (resp. $\Psi^t_{\si}$) 
and are given by the expressions (let $\s(x)\equiv 2\sin(\pi bx)$, 
$z=e^{2\pi b x}$)
\begin{equation}\begin{aligned}
\CBls{+}{\al_3}{\al_2}{\bar{\al}_4}{\al_1}{\Psi}(x)=& 
R(x)\;\bigl([2\al_1-b]   +
[\al_1+\al_4-\al_3-\fr{b}{2}]z\bigr)\\
\CBls{-}{\al_3}{\al_2}{\bar{\al}_4}{\al_1}{\Psi}(x)=&
R(x)\;[\al_4+\al_3+\al_1-\fr{3b}{2}-b^{-1}]z\\
\CBlt{+}{\al_3}{\al_2}{\bar{\al}_4}{\al_1}{\Psi}(x)=& 
R(x)\;\bigl([2\al_3-b] z  +
[\al_3+\al_4-\al_1-\fr{b}{2}] \bigr)\\
\CBls{-}{\al_3}{\al_2}{\bar{\al}_4}{\al_1}{\Psi}(x)=&
R(x)\;[\al_4+\al_3+\al_1-\fr{3b}{2}-b^{-1}],
\end{aligned}\end{equation}
where $R(x)$ abbreviates the common factor that appears. It now follows 
easily that the matrix $G_{s,t}$ introduced in Section 3 coincides
with the Racah coefficients that relate  $\Psi^s_{\si}$ and $\Psi^t_{\si}$:
\begin{equation} G_{s,t}(\al_4,\al_3,\al_1)\equiv
 \SJS{\al_1}{\al_2}
{\al_3}{\al_4}{\al_1-s\al_2}{\al_3-t\al_2}\Bigl|_{\al_2=-\frac{b}{2}},
 \qquad\text{where}\quad s,t=+,-.
\end{equation}
But this already guarantees that the finite difference equations 
that follow from the Moore-Seiberg pentagon equation for the Racah
coefficients as sketched in Section 3 will be exactly the same as
those satisfied by the fusion coefficients 
$G_{\al_{21},\al_{32}}\bigl[{}^{\al_3}_{\al_4}{}^{\al_2}_{\al_1}\bigr]$.

One can furthermore show that the functional equations that describe
the relation between Racah coefficients with argument $\al_i$ and $Q-\al_i$,
$i=1,\ldots,4$ coincide with those satisfied by 
$G_{\al_{21},\al_{32}}\bigl[{}^{\al_3}_{\al_4}{}^{\al_2}_{\al_1}\bigr]$.

We conclude that the Racah coefficients solve the full system of functional 
equations for the fusion coefficients.
This implies according to Section 3 that one has at least for 
real, irrational values of $b$ 
\begin{equation}
\SJSL{\al_1}{\al_2}{\al_3}{\al_4}{\al_{21}}{\al_{32}}=
\SJS{\al_1}{\al_2}{\al_3}{\al_4}{\al_{21}}{\al_{32}}.
\end{equation}

\section{Discussion}
The results of the present paper amount to a verification of consistency
of the bootstrap for Liouville theory under the assumptions on existence
and properties of fusion coefficients discussed in Sections 2 and 3.

It should be emphasized that characterizing the fusion coeffcients 
as Racah-coefficients for a quantum group will be important beyond
the task of veryifying the consistency of previous results on 
three point function and spectrum of Liouville theory. 
It also allows one
to complete the bootstrap for the associated boundary problem 
(Liouville theory on the strip or half-plane) that was begun in 
\cite{TZ} and should have important applications to D-brane physics
on the associated noncompact backgrounds. Specifically, it was observed
by one of us (J.T.) more than two years ago that the three point function
for {\it boundary} operators can be expressed in terms of the fusion 
coefficients, which was one of the main motivations for undertaking 
the present investigation. 

In fact, due to the close relationship between Liouville theory and the 
$H_3^+$ or 
$SL(2)/U(1)$ WZNW models it should now not be too difficult to generalize
our methods to obtain similar results for these models. 
For example,
an exact investigation of effective field theories on D-branes in $ADS_3$
similar to what was done in \cite{ARS,FFFS} is now within reach.

Our results may also be interesting from the mathematical point of 
view. The list of examples for non-compact quantum groups where
results on the harmonic analysis are known is rather short
\cite{W3,Ka,BR}. 
In fact, 
the ``deformation'' of classical groups to quantum groups often meets
subtle obstacles \cite{W1}. Here we have found an example which 
is {\it not} a deformation of a classical group but in  
some respects looks particularly nice (self-duality).

We would finally like to mention that there seem to be rather interesting
connections of the present work to other approaches to Liouville theory,
namely the Liouville model on the lattice \cite{FKV} and 
quantization of Teichm\"{u}ller space \cite{F1, K}. In fact, in all cases
the special function $S_b(x)$ (called quantum dilogarithm there) 
as well as the duality $b\ra b^{-1}$ play crucial roles. 
Making contact with the quantization of Teichm\"{u}ller space \cite{F1, K}
will require diagonalization of finite difference operators of similar 
form as have appeared in the present work \cite{F2}.

\section{Appendix: Special functions}
The basic building block for the class of special functions to be considered
is the the Double Gamma function introduced by Barnes \cite{Ba},
see also \cite{Sh}.
The Double Gamma function is defined as
\begin{equation}
\log\Ga_2(s|\om_1,\om_2)=  \Biggl(\frac{\pa}{\pa t}\sum_{n_1,n_2=0}^{\infty}
(s+n_1\om_1+n_2\om_2)^{-t}\Biggr)_{t=0}.
\end{equation}
Let $\Ga_b(x)=\Ga_2(x|b,b^{-1})$, and define 
the Double Sine function $S_b(x)$ and the Upsilon function $\up_b(x)$
respectively by
\begin{equation}
S_b(x)=\frac{\Ga_b(x)}{\Ga_b(Q-x)}\qquad \up_b(x)=\Ga_b(x)\Ga_b(Q-x).
\end{equation}
It will also be useful to introduce
\begin{equation}
G_b(x)=e^{\frac{\pi i}{2}x(x-Q)}S_b(x).
\end{equation}

The b-hypergeometric function will be defined by an integral representation
that resembles the Barnes integral for the ordinary hypergeometric function:
\begin{equation}\label{Barnesint}
F_b(\al,\be;\ga;y)=\frac{1}{i}\frac{S_b(\ga)}{S_b(\al)S_b(\be)}
\int_{-i\infty}^{i\infty}ds\;\, e^{2\pi i sy}
\frac{S_b(\al+s)S_b(\be+s)}{S_b(\ga+s)
S_b(Q+s) },
\end{equation}
where the contour is to the right of the poles at $s=-\al-nb-mb^{-1}$
$s=-\be-nb-mb^{-1}$ and to the left of the poles at $s=nb+mb^{-1}$
$s=Q-\ga+nb+mb^{-1}$, $n,m=0,1,2,\ldots$.
The function $F_b(\al,\be;\ga;-ix)$ is a solution of the $q$-hypergeometric
difference equation
\begin{equation}
\bigl([\de_x+\al][\de_x+\be]-e^{-2\pi b x}[\de_x][\de_x+\ga-Q]\bigr)
F_b(\al,\be;\ga;-ix)=0, \qquad \de_x=\fr{1}{2\pi}\pa_x
\end{equation}
This definition of a q-hypergeometric function is closely related to
the one first given in \cite{NU}.
There is also a kind of deformed Euler-integral 
for the hypergeometric function \cite{NU}:
\begin{equation}\label{Eulerint}
\Psi_b(\al,\be;\ga;x)=\frac{1}{i}\int_{-i\infty}^{i\infty}ds \;\,
e^{2\pi i s \be}\frac{G_b(s+x)G_b(s+\ga-\be)}{G_b(s+x+\al)G_b(s+Q)}
\end{equation}
The precise relation between $\Psi_b$ and $F_b$ is
\begin{equation}
\Psi_b(\al,\be;\ga;x)=\frac{G_b(\be)G_b(\ga-\be)}{G_b(\ga)}
F_b(\al,\be;\ga;x'),\qquad x'=x-\fr{1}{2}(\ga-\al-\be+Q).
\end{equation}
\newcommand{\CMP}[3]{{\it Comm. Math. Phys. }{\bf #1} (#2) #3}
\newcommand{\LMP}[3]{{\it Lett. Math. Phys. }{\bf #1} (#2) #3}
\newcommand{\IMP}[3]{{\it Int. J. Mod. Phys. }{\bf A#1} (#2) #3}
\newcommand{\NP}[3]{{\it Nucl. Phys. }{\bf B#1} (#2) #3}
\newcommand{\PL}[3]{{\it Phys. Lett. }{\bf B#1} (#2) #3}
\newcommand{\MPL}[3]{{\it Mod. Phys. Lett. }{\bf A#1} (#2) #3}
\newcommand{\PRL}[3]{{\it Phys. Rev. Lett. }{\bf #1} (#2) #3}
\newcommand{\AP}[3]{{\it Ann. Phys. (N.Y.) }{\bf #1} (#2) #3}
\newcommand{\LMJ}[3]{{\it Leningrad Math. J. }{\bf #1} (#2) #3}
\newcommand{\FAA}[3]{{\it Funct. Anal. Appl. }{\bf #1} (#2) #3}
\newcommand{\PTPS}[3]{{\it Progr. Theor. Phys. Suppl. }{\bf #1} (#2) #3}
\newcommand{\LMN}[3]{{\it Lecture Notes in Mathematics }{\bf #1} (#2) #2}

\end{document}